\documentclass[conference]{IEEEtran}
\IEEEoverridecommandlockouts
\usepackage{cite}
\usepackage{amsmath,amssymb,amsfonts}
\usepackage{algorithmic}
\usepackage{graphicx}
\usepackage{textcomp}
\usepackage{xcolor}
\def\BibTeX{{\rm B\kern-.05em{\sc i\kern-.025em b}\kern-.08em
    T\kern-.1667em\lower.7ex\hbox{E}\kern-.125emX}}
\begin{document}

\title{Evaluating the Effectiveness of Large Language Models in Solving Simple Programming Tasks: A User-Centered Study
}

\author{\IEEEauthorblockN{Kai Deng}
\IEEEauthorblockA{
\textit{Pine View School}\\
Osprey, FL, USA \\
yingkai.deng8@gmail.com}
}

\maketitle

\begin{abstract}
As large language models (LLMs) become more common in educational tools and programming environments, questions arise about how these systems should interact with users. This study investigates how different interaction styles with ChatGPT-4o (passive, proactive, and collaborative) affect user performance on simple programming tasks. I conducted a within-subjects experiment where fifteen high school students participated, completing three problems under three distinct versions of the model. Each version was designed to represent a specific style of AI support: responding only when asked, offering suggestions automatically, or engaging the user in back-and-forth dialogue.Quantitative analysis revealed that the collaborative interaction style significantly improved task completion time compared to the passive and proactive conditions. Participants also reported higher satisfaction and perceived helpfulness when working with the collaborative version. These findings suggest that the way an LLM communicates, how it guides, prompts, and responds, can meaningfully impact learning and performance.This research highlights the importance of designing LLMs that go beyond functional correctness to support more interactive, adaptive, and user-centered experiences, especially for novice programmers.
\end{abstract}

\begin{IEEEkeywords}
Large Language Models (LLMs), AI-Assisted Programming, Interaction Style, User-Centered Evaluation
\end{IEEEkeywords}

\section{Introduction}

Large language models (LLMs) are rapidly changing how people learn to code. Tools like ChatGPT\cite{achiam2023gpt}, trained on vast datasets of natural language and programming examples, can now explain concepts, generate code, answer questions, and even help debug—all through a conversational interface\cite{lin2025autop2cllmbasedagentframework,khosrawi2022conversational}. These models are increasingly used in educational settings\cite{ubiminds2024friend}, offering students on-demand support that feels more like a tutor than a textbook. As this shift continues, a key question arises: how should these tools interact with users to be most effective\cite{ubiminds2024friend,feng2025automation}? Most current research on LLMs focuses on whether having access to a tool like ChatGPT improves performance. These studies often treat the model as a single fixed entity, rather than exploring how different interaction styles might influence learning or user experience\cite{liu2024toursynbio,shen2024toursynbio,shen2024fine}. Yet, in human education, how support is given—whether through passive answers, proactive suggestions, or collaborative questioning—can make a significant difference. It stands to reason that the same may be true when the teacher is an AI \cite{luo2024autom3l}. This study was inspired by a simple question: if I were learning to code, what kind of AI interaction would actually help me improve? Would it be better if the AI waited for me to ask for help, offered suggestions automatically, or interacted with me like a partner in problem-solving? These are not just design questions—they are educational questions. As LLMs become more common in both classrooms and self-learning environments, understanding the style of support they provide is essential. To explore this, I designed a within-subjects experiment in which 15 high school students solved three programming tasks using three different versions of ChatGPT-4o. Each version represented a different interaction style: passive (help only when asked), proactive (automated suggestions), and collaborative (dialogue and shared reasoning). I collected both performance data and user feedback to evaluate not just which style worked best, but how it made participants feel. While previous research has shown that LLMs can support programming education\cite{yang2025effectiveness,zhang2024exploring}, this study focuses on how the form of that support affects both outcomes and experience—something that has been largely overlooked.

\subsection{Related Works}

Prior studies have explored how large language models, or LLMs, such as ChatGPT, influence the way students approach coding tasks. Sun et al. (2024)\cite{sun2024would}, for example, investigated how the presence of ChatGPT changed college students' behavior, confidence, and performance in programming assignments. While their research provided valuable insights into user perception and tool usage patterns, it did not include controlled comparisons across different types of interaction styles. Similarly, Khurana et al. (2024)\cite{khurana2024and} examined prompt effectiveness in help-seeking scenarios but focused more on technical outcomes than on the broader educational implications of interaction style. Other surveys, such as those by Sergeyuk et al. (2024)\cite{sergeyuk2025human} and Zuo et al. (2024)\cite{jin2024llms}, reviewed how LLMs integrate into software development and learning environments\cite{chen5047555chatcivic} but were largely theoretical or observational. What is missing in these studies is an experimental design that isolates how the form of LLM interaction—whether passive, proactive, or collaborative—affects actual user performance and perceived helpfulness\cite{shen2024proteinengine}. This study addresses that gap by directly comparing three distinct interaction models of ChatGPT in a controlled setting, with a focus on high school learners. In doing so, it contributes not only new data but also practical insights for designing AI tools that better support student needs\cite{amoozadeh2024student}.
\section{Materials and Methods}

This study used a within-subjects experimental design to investigate how different interaction styles with a large language model affect users’ ability to complete simple programming tasks. Each participant completed three tasks, each with a different version of ChatGPT-4o. The versions were designed to simulate three distinct interaction styles: a passive mode that only responded to direct requests, a proactive mode that suggested code or strategies without being asked, and a collaborative mode that engaged in back-and-forth dialogue to co-develop solutions. This setup allowed us to measure and compare performance across styles for each individual. Participants were high school students with a basic understanding of programming. Before beginning the experiment, each student completed a background survey to provide information on their prior coding experience, confidence level, and familiarity with AI tools. These responses confirmed that the group was relatively balanced in background knowledge, which helped ensure that differences in performance were more likely attributable to the interaction style than to pre-existing skill differences. Each participant was asked to solve three programming problems: one that involved adding two numbers represented as linked lists, another that required converting Roman numerals to integers, and a third that involved computing the square root of a number without using built-in math functions. These tasks were chosen because they reflect common beginner-level programming challenges that require algorithmic thinking but are straightforward to evaluate for correctness. The experiment was conducted remotely. All participants accessed the same task descriptions and input/output formats through a shared document. Those in the experimental condition received a custom link to a GPT interface tailored to one of the three interaction styles. Participants were instructed to record their code, note how long each task took, and complete a post-task questionnaire evaluating their experience. They were asked to work independently, and no external tools beyond the provided GPT version were allowed. To analyze the results, we collected three types of data: task completion time, correctness of submitted code, and participant feedback from the post-task survey. We conducted a Shapiro-Wilk test to confirm normality and used a repeated-measures ANOVA to evaluate whether GPT interaction style had a significant effect on task completion time. Tukey’s Honest Significant Difference test was used for post-hoc pairwise comparisons to determine which styles differed significantly from each other. Together, these measures provided both quantitative and qualitative insights into how interaction style shaped the coding process. 

\subsection{Model Specifications}

All assistance in this study was provided using ChatGPT-4o, a version of OpenAI’s large language model designed for fast, context-aware, multi-turn interactions. ChatGPT-4o was selected for its conversational flexibility, consistency across prompts, and accessibility to users at the time of the study. While this study used a specific implementation (ChatGPT-4o), the findings aim to inform broader discussions about how interaction styles may affect user performance when working with large language models (LLMs). 

\subsection{Hypotheses}

This study was guided by three main hypotheses. First, it was expected that the style of interaction with ChatGPT-4o—whether passive, proactive, or collaborative\cite{hu2025generation}—would significantly affect task completion time when solving programming problems. Second, it was hypothesized that participants would complete tasks more efficiently when interacting with the model in a collaborative style, as compared to passive or proactive styles. Third, it was anticipated that participants would report a more positive user experience when engaged in collaborative interactions. No specific magnitude of improvement was predicted in advance, given the exploratory nature of the study. 

\subsection{Quantitative Metrics}

Three quantitative metrics were used to evaluate the effectiveness of each ChatGPT-4o interaction style. Task completion time, measured in minutes, captured how long participants spent solving each programming task under each interaction condition. Code correctness was assessed by evaluating the submitted solutions against standard input and output expectations. Lastly, user feedback scores were collected through post-task surveys, which asked participants to rate their perceptions of helpfulness, confidence, and satisfaction using Likert-scale questions. 

\subsection{Procedure}

Participants completed the study individually, using shared digital documents and customized ChatGPT-4o links. For each task, participants received the task prompt and the appropriate assistant link. They were instructed to complete the tasks as accurately and efficiently as possible without external assistance beyond the materials provided. After completing all tasks, participants filled out a feedback questionnaire evaluating their experiences with each interaction style. 

\subsection{Data Collection and Analysis}

This study collected three primary types of data: task completion time, submitted code, and post-task feedback. Task time was measured in minutes to evaluate efficiency. Submitted code was reviewed for functional correctness and logical soundness. Post-task feedback captured user experience and perception through survey responses. To analyze the data, a Shapiro-Wilk test was first performed to verify the normality of the task time distributions. A repeated-measures ANOVA was then used to determine whether the ChatGPT-4o interaction style had a significant effect on task performance. When significant differences were found, Tukey HSD post-hoc tests were conducted to identify specific pairwise contrasts between the interaction styles.

\section{Results}

\subsection{Task Completion Time Analysis}

\begin{figure}[tb]
  \centering
  \resizebox{\linewidth}{!}{%
    \includegraphics{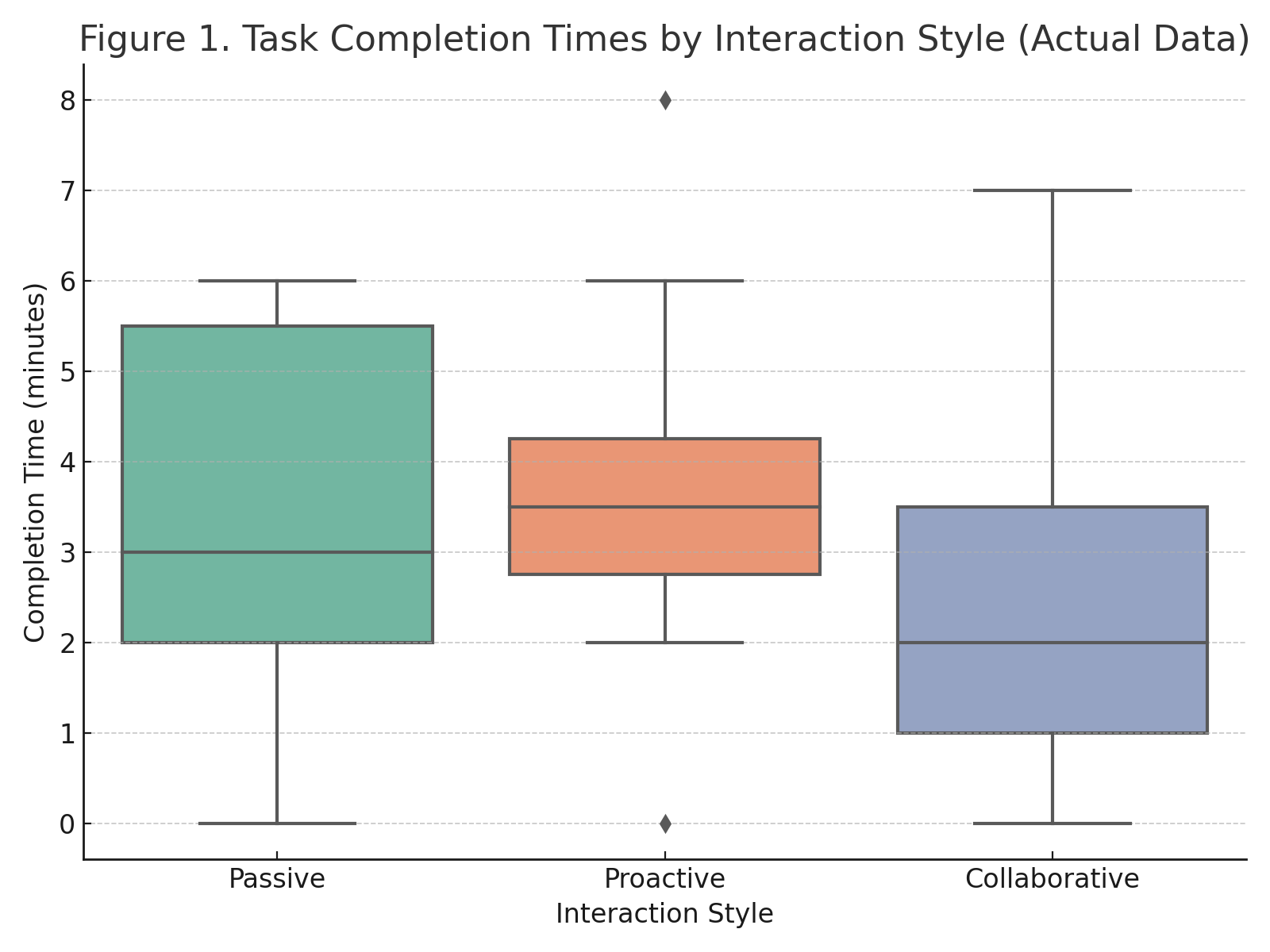}%
  }
  \caption{Boxplot of task completion times across three GPT interaction styles.}
  \label{fig:boxplot}
\end{figure}

\begin{figure}[tb]      
  \centering
  \resizebox{\linewidth}{!}{%
    \includegraphics{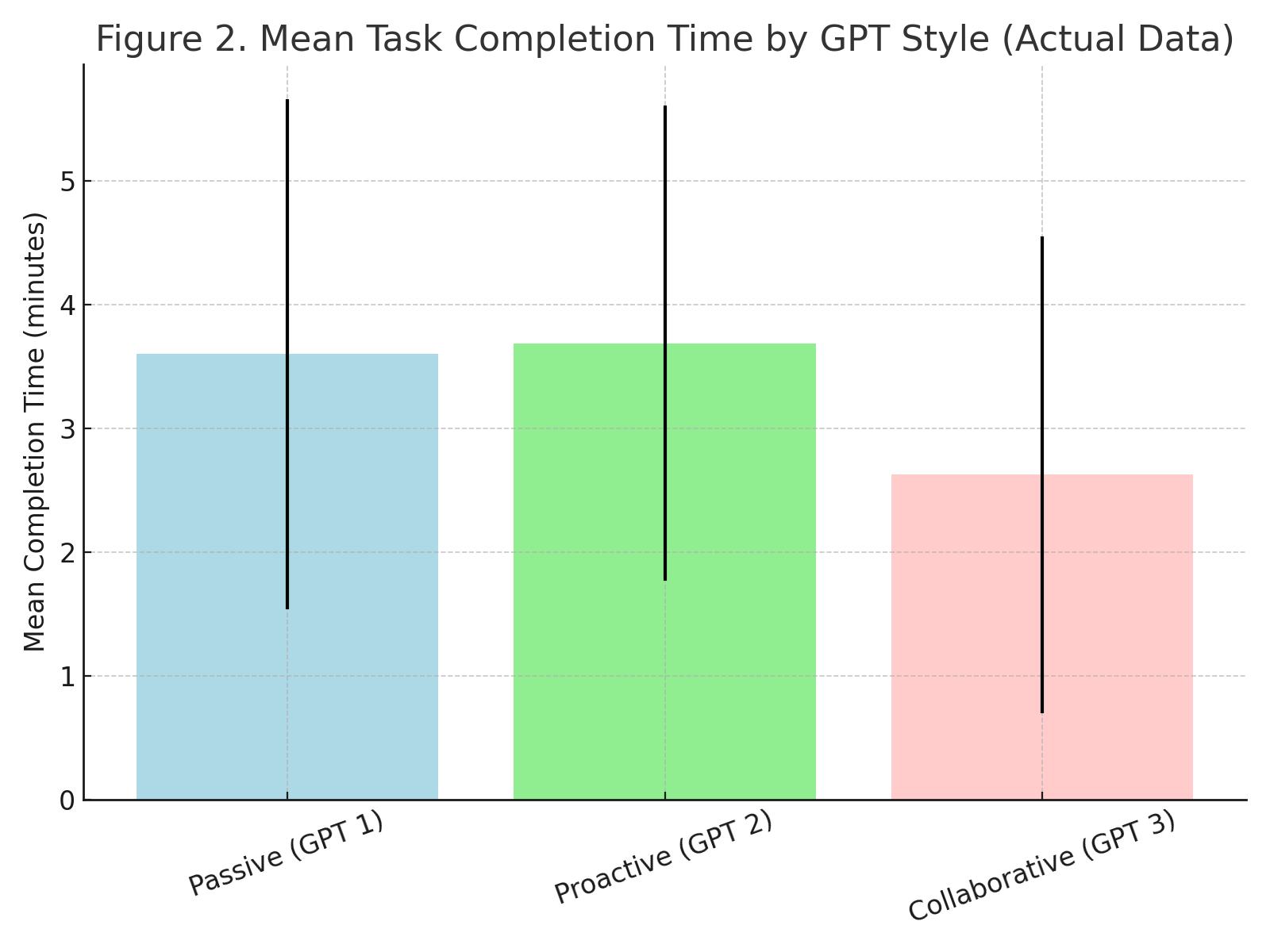}
  }
  \caption{Mean task-completion time by GPT style with standard-deviation error bars.}
  \label{fig:mean_time}    
\end{figure}

A repeated-measures ANOVA was conducted to assess whether interaction style (passive, proactive, collaborative) significantly affected task completion time. The results revealed a statistically significant main effect of interaction style, F (2, 28) = 5.35, p = 0.0108. Normality of task completion times was verified using the Shapiro-Wilk test for each condition (all p-values $> 0.18$), supporting the use of parametric analysis. Descriptive statistics for task completion times are presented in Table \ref{tab:time_stats}. The distribution of these times is visualized in Fig.~\ref{fig:boxplot}, and their means are compared in Fig.~\ref{fig:mean_time}.

\begin{table}[tb] 
  \caption{Descriptive Statistics for Task Completion Time Across Interaction Styles}
  \label{tab:time_stats}
  \centering
  \resizebox{\linewidth}{!}{%
    \begin{tabular}{lcc}
      \hline
      \textbf{Interaction Style} &
      \textbf{Mean Completion Time (min)} &
      \textbf{Standard Deviation (min)} \\
      \hline
      Passive (Task~1)       & 3.60 & 2.06 \\
      Proactive (Task~2)     & 3.69 & 1.92 \\
      Collaborative (Task~3) & 2.63 & 1.93 \\
      \hline
    \end{tabular}
  }
\end{table}

Post-hoc pairwise comparisons using Tukey's HSD test revealed several key differences between interaction styles. Participants completed tasks significantly faster when using the Collaborative style compared to the Passive style, with a p-value of 0.004. A statistically significant improvement was also observed when comparing the Proactive style to the Passive style, with a p-value of 0.047. However, no significant difference was found between the Collaborative and Proactive styles, as indicated by a p-value of 0.489. 

\subsection{Post-Task Survey Feedback}

\begin{figure}[tb]                
  \centering
  \resizebox{\linewidth}{!}{
    \includegraphics{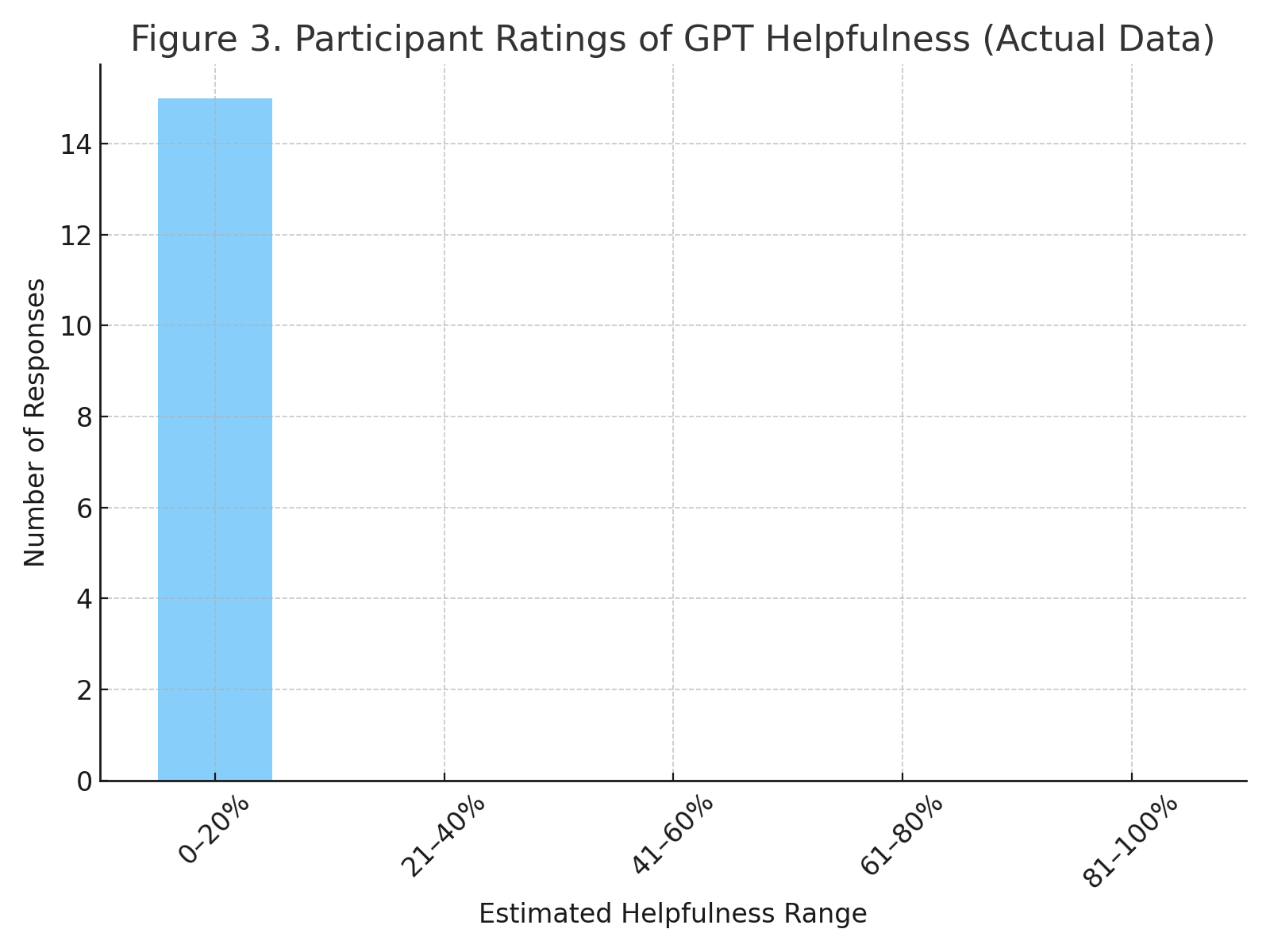}
  }
  \caption{Participant responses to the question “How helpful was GPT during the coding tasks?”}
  \label{fig:helpfulness}
\end{figure}

Survey responses collected after the tasks offered additional insight into participants’ experiences with the different interaction styles. Most participants found ChatGPT-4o easy or very easy to interact with, and the majority rated its suggestions as clear or very clear. In terms of satisfaction and productivity, participants generally reported feeling satisfied or very satisfied with the assistance they received, with many expressing that it helped them work more quickly and efficiently. Participant ratings on the helpfulness of the AI are shown in Fig.~\ref{fig:helpfulness}. Feedback on confidence was more mixed; while some participants indicated that the AI improved their confidence, others reported no change or even a decrease, suggesting that the impact of interaction style on self-perception may vary depending on individual learning preferences. When asked to estimate the proportion of directly helpful code, participants reported that approximately 6 to 10 percent of the output provided by ChatGPT-4o contributed meaningfully to their final solutions. 

These results suggest that the style of interaction with a large language model meaningfully impacts task efficiency and user perceptions during programming tasks. Collaborative-style assistance led to faster completion times and generally higher satisfaction compared to more passive forms of support. In the following section, we discuss the implications of these findings, how they relate to previous work, and what they suggest about the future design of AI-assisted learning environments. 
\section{Discussion}

\subsection{Summary of Findings}

This study was aimed at determining whether the style of interaction with a large language model, specifically ChatGPT-4o, impacted the effectiveness and experience of solving simple programming problems. The results confirmed that interaction style does matter. Participants using a collaborative GPT interface completed tasks significantly faster than those using a passive or proactive version. While the proactive model offered some improvement over the passive version, the gains were not statistically significant compared to the collaborative model. Post-task surveys also revealed higher satisfaction and clarity ratings in the collaborative group, reinforcing the performance results with subjective feedback. 

\subsection{Interpretation and Implications}

Our findings indicate that the quality of human-AI interaction, rather than merely the accuracy of the model's output, is a critical factor in the educational utility of LLMs. The collaborative GPT didn't just wait to be prompted (like the passive version) or blindly offer suggestions (like the proactive one\cite{sharma2024carry}). Instead, it engaged the user in a more natural, back-and-forth coding dialogue\cite{allen1999mixed}. This seemed to help participants stay oriented, build momentum, and feel supported throughout the task. Rather than simply speeding up the process, this form of interaction reshaped the experience of programming itself. In contrast, the passive GPT may have required too much initiative from users, many of whom were still learning how to ask for help effectively\cite{7827078}. The proactive version offered more assistance but lacked the adaptive, conversational flow of the collaborative model. These distinctions show that LLMs can't just be powerful, they also need to be psychologically attuned to how users think and work, especially in a learning context\cite{schulz2024generative}. 

\subsection{Connection to Prior Work }

These results align with and build upon the findings of Sun et al. (2024)\cite{sun2024would}, who observed that students using ChatGPT during programming assignments demonstrated changes in behavior, including increased engagement and improved perceptions of their own abilities. However, their study focused primarily on how students use ChatGPT in authentic classroom settings, without directly comparing performance across different AI interaction designs. Our controlled setup extends this line of inquiry by showing that not all GPT usage is equal\cite{klieger2024chatcollab}—the way a model engages with the user significantly affects outcomes. Other recent studies in human-AI interaction (e.g., \cite{khurana2024and}) have emphasized the importance of contextual, adaptive AI behaviors in support tools\cite{richards2025bridging,ma2023should}. These studies often point to the need for AI systems that can understand user intent, adjust their level of support, and avoid overwhelming or under-supporting the user\cite{vasconcelos2023explanations}. The collaborative GPT model used in our experiment reflects many of these principles: rather than acting like a search engine or code generator, it behaved more like a peer—asking questions, checking in, and helping steer the user toward a solution\cite{shi2025reinforcementfinetuningreasoningmultistep,yu2025co}. This distinction is especially critical for novice programmers, who may struggle not only with technical syntax but with the confidence to keep going when they get stuck\cite{ma2025dbox}. While prior research often centers on professionals or advanced students, our study contributes new insight into how AI tools can be optimized for beginner learning contexts, where emotional experience and cognitive scaffolding play a large role in success. 

\subsection{Educational Implications}

The implications of these findings extend beyond simple task performance. As large language models continue to be integrated into classrooms and self-learning platforms, educators and designers must consider how interaction style shapes learning outcomes. Our results suggest that a collaborative approach to AI guidance, one that mimics the flow of human tutoring, may be particularly effective for supporting novice coders\cite{cai2025exploring}. For educators, this means that simply giving students access to an LLM isn’t enough. The way the tool is presented, and the nature of its engagement may determine whether students benefit or become overwhelmed. Embedding LLMs into curricula with structured prompts, interactive scaffolding, and boundaries around over-reliance may foster more positive experiences. Especially for high school students and early learners, a model that encourages dialogue and questions may feel less intimidating and more empowering. For tool designers, the takeaway is equally clear: building smarter AI isn't just about improving accuracy or adding features. It's about designing socially aware models that can interpret user hesitation, respond flexibly, and encourage productive exploration. As this study shows, even relatively small design shifts, from passive to collaborative, can yield meaningful differences in outcomes. 

\subsection{Limitations}

While this study offers valuable insight into how GPT interaction styles influence programming task outcomes, several limitations should be acknowledged\cite{kapania2025m}. First, the sample size was relatively small (n = 15), and all participants were high school students with basic programming experience. This demographic provided a consistent user base for comparison, but it limits the generalizability of the findings. More diverse populations, including adults, university-level learners, or professional developers, may interact differently with large language models. Second, the tasks used in this study were intentionally simple. While this allowed for easier comparison and reduced cognitive overload, it also means the results may not extend to more complex, multi-step programming assignments or projects that require sustained debugging over time. Third, the use of three pre-designed GPT variants (passive, proactive, collaborative) introduces some artificiality. In a real-world setting, users might switch styles organically or adjust their interaction habits over time. Our study isolates the effect of interaction style, but future research could explore how these styles evolve when users are allowed more freedom and agency. Finally, the study was conducted remotely, which introduced variability in participants' environments\cite{kosch2024risk}. Although we controlled the tools and materials used, factors such as distractions, internet stability, or motivation could have influenced results. 

\subsection{Future Work}

There are several ways this study could be expanded. First, future research should include a larger and more diverse group of participants. This would help determine whether the findings apply to other groups, like college students, adults learning to code, or even professional developers, who might interact with LLMs differently than high school students. Another direction would be to use a wider range of programming problems. The tasks in this study were simple and designed for beginners. It would be useful to see whether the same patterns hold when participants face more complex problems that require longer-term planning or debugging\cite{wang2024evaluating}. It’s possible that collaborative GPT support becomes even more useful in those cases. Another direction would be to test a version of GPT that adapts its behavior based on the user's needs, offering more help when someone is stuck, and stepping back when they're making progress. This could create a more personalized experience and help avoid giving too much or too little help. Finally, future studies could focus on long-term effects. Instead of just measuring how fast someone completes a task, researchers could track whether users actually improve over time, feel more confident, or retain what they learned\cite{wandb2024llmeval,wang2024evaluating,humanloop2025tools}. This would help us understand whether GPT-style support is helping people become better problem-solvers, not just faster ones.

\section{Conclusions}

This study explored how different interaction styles with a large language model, specifically ChatGPT-4o, influence users’ ability to solve simple programming problems. By comparing passive, proactive, and collaborative forms of AI assistance, we found that the collaborative style consistently led to faster task completion and higher user satisfaction. These results show that how an AI assistant interacts with users can be just as important as the information it provides. As large language models become more common in learning environments, this research highlights the need for systems that offer more than just correct answers. Models that guide users through an interactive experience, provide step-by-step support, and encourage problem solving may have a much stronger impact on learning than models that simply respond to prompts. Although this study focused on short tasks and beginner programmers, the findings raise broader questions about the future of AI in education. To build tools that support long-term learning and confidence, we may need to prioritize thoughtful interaction design as much as technical capability.

\bibliographystyle{IEEEtran}
\bibliography{main}

\begin{thebibliography}{10}
\providecommand{\url}[1]{#1}
\csname url@samestyle\endcsname
\providecommand{\newblock}{\relax}
\providecommand{\bibinfo}[2]{#2}
\providecommand{\BIBentrySTDinterwordspacing}{\spaceskip=0pt\relax}
\providecommand{\BIBentryALTinterwordstretchfactor}{4}
\providecommand{\BIBentryALTinterwordspacing}{\spaceskip=\fontdimen2\font plus
\BIBentryALTinterwordstretchfactor\fontdimen3\font minus \fontdimen4\font\relax}
\providecommand{\BIBforeignlanguage}[2]{{%
\expandafter\ifx\csname l@#1\endcsname\relax
\typeout{** WARNING: IEEEtran.bst: No hyphenation pattern has been}%
\typeout{** loaded for the language `#1'. Using the pattern for}%
\typeout{** the default language instead.}%
\else
\language=\csname l@#1\endcsname
\fi
#2}}
\providecommand{\BIBdecl}{\relax}
\BIBdecl

\bibitem{achiam2023gpt}
J.~Achiam, S.~Adler \emph{et~al.}, ``Gpt-4 technical report,'' \emph{arXiv preprint arXiv:2303.08774}, 2023.

\bibitem{lin2025autop2cllmbasedagentframework}
\BIBentryALTinterwordspacing
Z.~Lin, Y.~Shen, Q.~Cai, H.~Sun, J.~Zhou, and M.~Xiao, ``Autop2c: An llm-based agent framework for code repository generation from multimodal content in academic papers,'' 2025. [Online]. Available: \url{https://arxiv.org/abs/2504.20115}
\BIBentrySTDinterwordspacing

\bibitem{khosrawi2022conversational}
B.~Khosrawi-Rad, H.~Rinn, R.~Schlimbach, P.~Gebbing, X.~Yang, C.~Lattemann, D.~Markgraf, and S.~Robra-Bissantz, ``Conversational agents in education--a systematic literature review,'' 2022.

\bibitem{ubiminds2024friend}
Ubiminds, ``{AI-Assisted Coding: Friend or Foe?}'' \url{https://ubiminds.com/en-us/ai-assisted-coding/}, 2024.

\bibitem{feng2025automation}
T.~H. Feng, A.~Luxton-Reilly, B.~C. W{\"u}nsche, and P.~Denny, ``From automation to cognition: Redefining the roles of educators and generative ai in computing education,'' in \emph{Proceedings of the 27th Australasian Computing Education Conference}, 2025, pp. 164--171.

\bibitem{liu2024toursynbio}
Y.~Liu, Z.~Chen, Y.~G. Wang, and Y.~Shen, ``Toursynbio-search: A large language model driven agent framework for unified search method for protein engineering,'' in \emph{2024 IEEE International Conference on Bioinformatics and Biomedicine (BIBM)}.\hskip 1em plus 0.5em minus 0.4em\relax IEEE, 2024, pp. 5395--5400.

\bibitem{shen2024toursynbio}
Y.~Shen, Z.~Chen, M.~Mamalakis, Y.~Liu, T.~Li, Y.~Su, J.~He, P.~Li{\`o}, and Y.~G. Wang, ``Toursynbio: A multi-modal large model and agent framework to bridge text and protein sequences for protein engineering,'' in \emph{2024 IEEE International Conference on Bioinformatics and Biomedicine (BIBM)}.\hskip 1em plus 0.5em minus 0.4em\relax IEEE, 2024, pp. 2382--2389.

\bibitem{shen2024fine}
Y.~Shen, Z.~Chen, M.~Mamalakis, L.~He, H.~Xia, T.~Li, Y.~Su, J.~He, and Y.~G. Wang, ``A fine-tuning dataset and benchmark for large language models for protein understanding,'' in \emph{2024 IEEE International Conference on Bioinformatics and Biomedicine (BIBM)}.\hskip 1em plus 0.5em minus 0.4em\relax IEEE, 2024, pp. 2390--2395.

\bibitem{luo2024autom3l}
D.~Luo, C.~Feng, Y.~Nong, and Y.~Shen, ``Autom3l: An automated multimodal machine learning framework with large language models,'' in \emph{Proceedings of the 32nd ACM International Conference on Multimedia}, 2024, pp. 8586--8594.

\bibitem{yang2025effectiveness}
T.-C. Yang, Y.-C. Hsu, and J.-Y. Wu, ``The effectiveness of chatgpt in assisting high school students in programming learning: evidence from a quasi-experimental research,'' \emph{Interactive Learning Environments}, pp. 1--18, 2025.

\bibitem{zhang2024exploring}
L.~Zhang, J.~Shu, J.~Hu, F.~Li, J.~He, P.~Wang, and Y.~Shen, ``Exploring the potential of large language models in radiological imaging systems: improving user interface design and functional capabilities,'' \emph{Electronics}, vol.~13, no.~11, p. 2002, 2024.

\bibitem{sun2024would}
D.~Sun, A.~Boudouaia, C.~Zhu, and Y.~Li, ``Would chatgpt-facilitated programming mode impact college students’ programming behaviors, performances, and perceptions? an empirical study,'' \emph{International Journal of Educational Technology in Higher Education}, vol.~21, no.~1, p.~14, 2024.

\bibitem{khurana2024and}
A.~Khurana, H.~Subramonyam, and P.~K. Chilana, ``Why and when llm-based assistants can go wrong: Investigating the effectiveness of prompt-based interactions for software help-seeking,'' in \emph{Proceedings of the 29th International Conference on Intelligent User Interfaces}, 2024, pp. 288--303.

\bibitem{sergeyuk2025human}
A.~Sergeyuk, I.~Zakharov, E.~Koshchenko, and M.~Izadi, ``Human-ai experience in integrated development environments: A systematic literature review,'' \emph{arXiv preprint arXiv:2503.06195}, 2025.

\bibitem{jin2024llms}
H.~Jin, L.~Huang, H.~Cai, J.~Yan, B.~Li, and H.~Chen, ``From llms to llm-based agents for software engineering: A survey of current, challenges and future,'' \emph{arXiv preprint arXiv:2408.02479}, 2024.

\bibitem{chen5047555chatcivic}
J.~Chen and Y.~Bao, ``Chatcivic: A domain-specific large language model (llm) for design code interpretation,'' \emph{Available at SSRN 5047555}, 2024.

\bibitem{shen2024proteinengine}
Y.~Shen, O.~Lv, H.~Zhu, and Y.~G. Wang, ``Proteinengine: Empower llm with domain knowledge for protein engineering,'' in \emph{International Conference on Artificial Intelligence in Medicine}.\hskip 1em plus 0.5em minus 0.4em\relax Springer, 2024, pp. 373--383.

\bibitem{amoozadeh2024student}
M.~Amoozadeh, D.~Nam, D.~Prol, A.~Alfageeh, J.~Prather, M.~Hilton, S.~Srinivasa~Ragavan, and A.~Alipour, ``Student-ai interaction: A case study of cs1 students,'' in \emph{Proceedings of the 24th Koli Calling International Conference on Computing Education Research}, 2024, pp. 1--13.

\bibitem{hu2025generation}
T.~Hu and S.~Wang, ``From generation to adaptation: Comparing ai-assisted strategies in high school programming education,'' \emph{arXiv preprint arXiv:2506.15955}, 2025.

\bibitem{sharma2024carry}
K.~Sharma and M.~Giannakos, ``Carry-forward effect: providing proactive scaffolding to learning processes,'' \emph{Behaviour \& Information Technology}, pp. 1--40, 2024.

\bibitem{allen1999mixed}
J.~E. Allen, C.~I. Guinn, and E.~Horvtz, ``Mixed-initiative interaction,'' \emph{IEEE Intelligent Systems and their Applications}, vol.~14, no.~5, pp. 14--23, 1999.

\bibitem{7827078}
G.~Salvaneschi, S.~Proksch, S.~Amann, S.~Nadi, and M.~Mezini, ``On the positive effect of reactive programming on software comprehension: An empirical study,'' \emph{IEEE Transactions on Software Engineering}, vol.~43, no.~12, pp. 1125--1143, 2017.

\bibitem{schulz2024generative}
T.~Schulz, M.~T. Knierim, and C.~Weinhardt, ``How generative-ai-assistance impacts cognitive load during knowledge work: a study proposal,'' in \emph{NeuroIS Retreat}.\hskip 1em plus 0.5em minus 0.4em\relax Springer, 2024, pp. 357--365.

\bibitem{klieger2024chatcollab}
B.~Klieger, C.~Charitsis, M.~Suzara, S.~Wang, N.~Haber, and J.~C. Mitchell, ``Chatcollab: Exploring collaboration between humans and ai agents in software teams,'' \emph{arXiv preprint arXiv:2412.01992}, 2024.

\bibitem{richards2025bridging}
J.~Richards and M.~Wessel, ``Bridging hci and ai research for the evaluation of conversational se assistants,'' \emph{arXiv preprint arXiv:2502.07956}, 2025.

\bibitem{ma2023should}
S.~Ma, Y.~Lei, X.~Wang, C.~Zheng, C.~Shi, M.~Yin, and X.~Ma, ``Who should i trust: Ai or myself? leveraging human and ai correctness likelihood to promote appropriate trust in ai-assisted decision-making,'' in \emph{Proceedings of the 2023 CHI Conference on Human Factors in Computing Systems}, 2023, pp. 1--19.

\bibitem{vasconcelos2023explanations}
H.~Vasconcelos, M.~J{\"o}rke, M.~Grunde-McLaughlin, T.~Gerstenberg, M.~S. Bernstein, and R.~Krishna, ``Explanations can reduce overreliance on ai systems during decision-making,'' \emph{Proceedings of the ACM on Human-Computer Interaction}, vol.~7, no. CSCW1, pp. 1--38, 2023.

\bibitem{shi2025reinforcementfinetuningreasoningmultistep}
\BIBentryALTinterwordspacing
W.~Shi and Y.~Shen, ``Reinforcement fine-tuning for reasoning towards multi-step multi-source search in large language models,'' 2025. [Online]. Available: \url{https://arxiv.org/abs/2506.08352}
\BIBentrySTDinterwordspacing

\bibitem{yu2025co}
J.~Yu, Y.~Wu, Y.~Zhan, W.~Guo, Z.~Xu, and R.~Lee, ``Co-learning: code learning for multi-agent reinforcement collaborative framework with conversational natural language interfaces,'' \emph{Frontiers in Artificial Intelligence}, vol.~8, p. 1431003, 2025.

\bibitem{ma2025dbox}
S.~Ma, J.~Wang, Y.~Zhang, X.~Ma, and A.~Y. Wang, ``Dbox: Scaffolding algorithmic programming learning through learner-llm co-decomposition,'' in \emph{Proceedings of the 2025 CHI Conference on Human Factors in Computing Systems}, 2025, pp. 1--20.

\bibitem{cai2025exploring}
L.~Cai, M.~M. Msafiri, and D.~Kangwa, ``Exploring the impact of integrating ai tools in higher education using the zone of proximal development,'' \emph{Education and Information Technologies}, vol.~30, no.~6, pp. 7191--7264, 2025.

\bibitem{kapania2025m}
S.~Kapania, R.~Wang, T.~J.-J. Li, T.~Li, and H.~Shen, ``'i'm categorizing llm as a productivity tool': Examining ethics of llm use in hci research practices,'' \emph{Proceedings of the ACM on Human-Computer Interaction}, vol.~9, no.~2, pp. 1--26, 2025.

\bibitem{kosch2024risk}
T.~Kosch and S.~Feger, ``Risk or chance? large language models and reproducibility in human-computer interaction research,'' \emph{arXiv e-prints}, pp. arXiv--2404, 2024.

\bibitem{wang2024evaluating}
L.~Wang and Y.~Shen, ``Evaluating causal reasoning capabilities of large language models: A systematic analysis across three scenarios,'' \emph{Electronics}, vol.~13, no.~23, p. 4584, 2024.

\bibitem{wandb2024llmeval}
WandB, ``{LLM Evaluations: Metrics, Frameworks, and Best Practices},'' \url{https://wandb.ai/onlineinference/genai-research/reports/LLM-evaluations-Metrics-frameworks-and-best-practices--VmlldzoxMTMxNjQ4NA}, 2024, accessed: 5 Jun 2025.

\bibitem{humanloop2025tools}
Humanloop, ``{5 LLM Evaluation Tools You Should Know in 2025},'' \url{https://humanloop.com/blog/best-llm-evaluation-tools}, 2025, accessed: 5 Jun 2025.

\end{thebibliography}

\end{document}